\documentclass[a4paper]{jpconf}
\usepackage{graphicx}
\begin{document}
\title{Growth Kinetics of the Homogeneously Nucleated Water Droplets:
Simulation Results}

\author{Anatolii V. Mokshin and Bulat N. Galimzyanov}

\address{Department of Physics, Kazan Federal University,
420008 Kazan, Russia}

\ead{anatolii.mokshin@mail.ru}

\begin{abstract}
The growth of homogeneously nucleated droplets in water vapor at
the fixed temperatures $T=273$, $283$, $293$, $303$, $313$, $323$,
$333$, $343$, $353$, $363$ and $373$~K (the pressure $p=1$~atm.)
is investigated on the basis of the coarse-grained molecular
dynamics simulation data with the mW-model. The treatment of
simulation results is performed by means of the statistical method
within the mean-first-passage-time approach, where the reaction
coordinate is associated with the largest droplet size. It is
found that the water droplet growth is characterized by the next
features: (i) the rescaled growth law is unified at all the
considered temperatures and (ii) the droplet growth evolves with
acceleration and follows the power law.
\end{abstract}

\section{Introduction}

Nucleation is one of the possible ways, by means of which a phase
transition can be initiated. According to the classical nucleation
theory, it is assumed that nucleus (cluster) of a new phase
becomes stable to grow in a `mother phase', when its size reaches
some critical value -- the critical size $n_c$. Consequently, both
related processes, nucleation and nuclei growth, belong to the
initial stages of phase transition. The character of phase
transition depends directly on the nucleation rate $J_s$ and the
growth rate $G$. Appearance of liquid droplets (bubbles) as
origins of condensation (vaporization) and emergence of
crystallite grains at crystallization are the typical examples of
nucleation. A feature of homogeneous
nucleation\footnote{Homogeneous scenario means the same
probability for the nucleation event emergence over the whole
system.} for most of molecular systems at the natural conditions
is that the characteristic spatial
scale~\cite{Kashchiev_Nucleation}, which is associated with the
linear critical size of the nucleated phase, takes the values
$10^{-9} \div 10^{-7}$m. Thereby, the processes of nucleation and
of nuclei growth are excellent candidates to be studied by means
of the methods of molecular dynamics simulation. However, the
probabilistic character of nucleation (the values of nucleation
rate, critical size, growth rate and others) prompts applying the
statistical treatment of the simulation results.

Recently, it was shown that the mean-first-passage-time (MFPT)
approach can be successfully used to define all the
characteristics of nucleation kinetics from the data of molecular
dynamics
simulations~\cite{Wedekind_JCP_2007,Mokshin/Barrat_JCP_2009}. In
the idealized case of the symmetric nucleation barrier, the
MFTP-distribution obtained from the set of independent simulations
or from experiments is interpolated in the nucleation regime by
the error-function, and such parameters as the steady-state
nucleation rate $J_s$ (or the nucleation  time scale $\tau_s=(J_s
V)^{-1}$, $V$ is the system volume), the critical size $n_c$ and
the curvature of the nucleation barrier associated with the
Zeldovich factor $Z$ can be extracted by numerical fitting. An
extended MFTP approach allows one to define all these parameters
for a wider class of nucleation scenario by numerical analysis of
the first derivative of the MFPT-curve with respect to the
reaction coordinate (say, nucleus size). In the present work we
show that the same MFPT-approach can be also applied to define the
characteristics of the nuclei growth regime. We demonstrate this
analysis as applied for the case of the water droplet growth.

\section{Methods and results}

\begin{figure}[h]
\begin{minipage}{10cm}
\includegraphics[width=10cm]{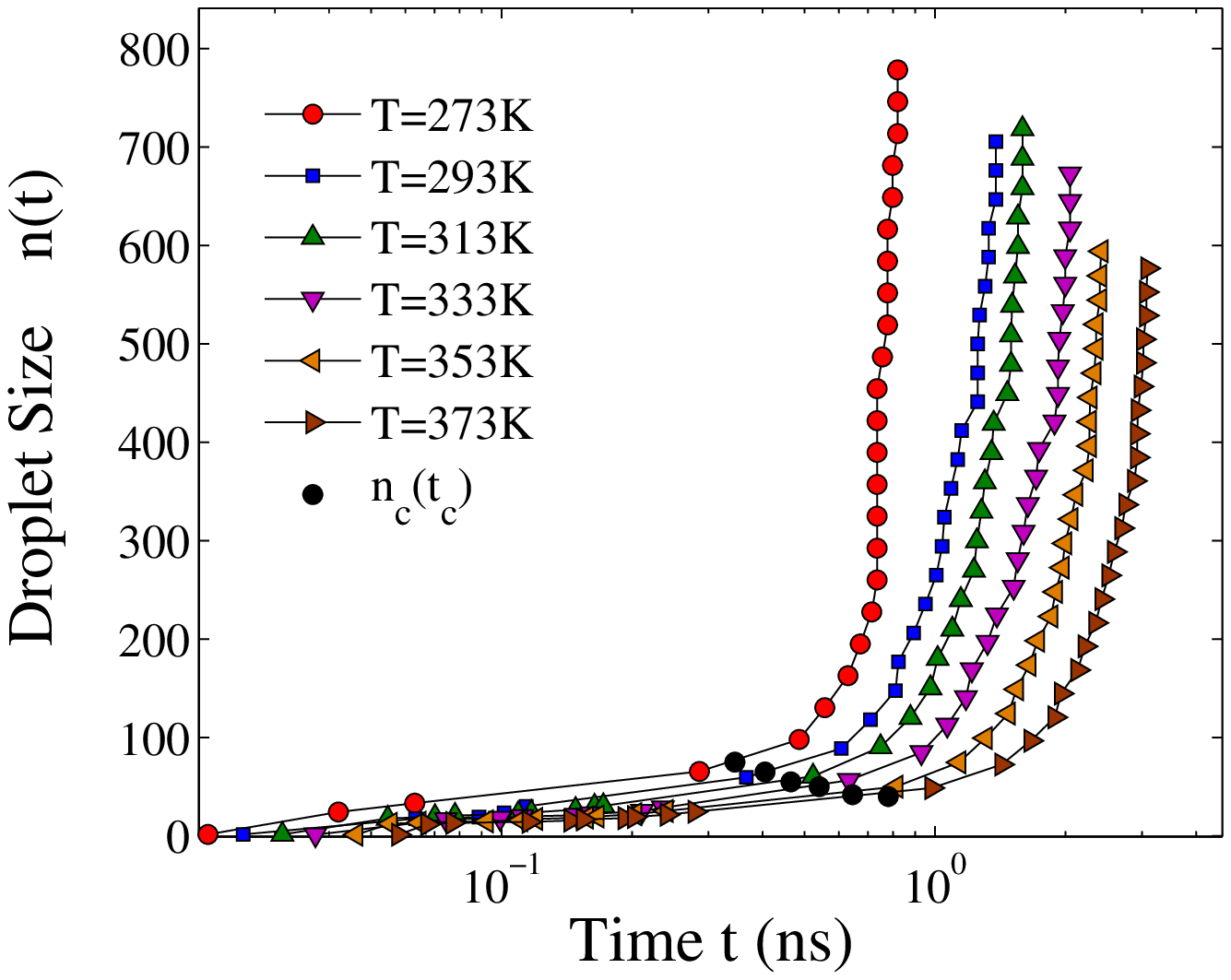}
\caption{\label{fig_growth_curves} Growth curves of the nucleated
water droplet as obtained from $NpT$-simulations within the
MFPT-approach. The pressure is the same for all the considered
cases, $p=1$~atm. Black circles indicate the critical sizes
$n_c$.}
\end{minipage}\hspace{2pc}%
\begin{minipage}{12pc}
\includegraphics[width=12pc]{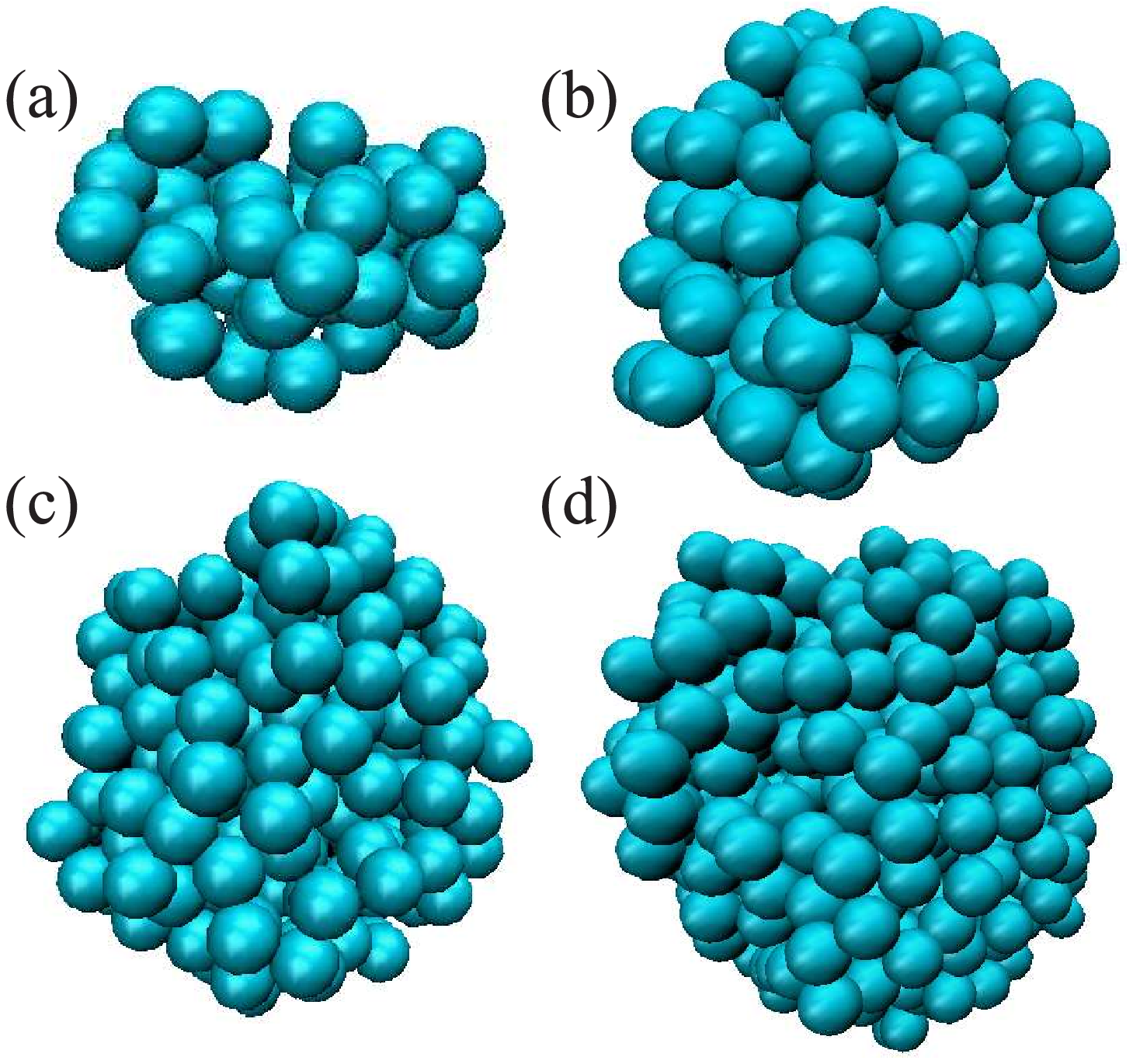}
\caption{\label{fig_snapshots of droplet} Water droplet at the
temperature $T=293$~K and at the different time after nucleation:
(a) $n=55$~molecules, $t(n)=0.75$~ns, (b) $n=123$~molecules,
$t(n)=0.83$~ns, (c) $n=197$~molecules, $t(n)=0.88$~ns, (d)
$n=308$~molecules, $t(n)=0.91$~ns. Circles are the conventional
symbols of water molecules.}
\end{minipage}
\end{figure}

%\begin{figure}[h]
%%\begin{center}
%\begin{minipage}{12pc}
%\includegraphics[width=12cm]{fig1.eps}
%\caption{\label{label}Figure caption for first of two sided
%figures.}
%\end{minipage}
%\begin{minipage}{12pc}
%\includegraphics[width=12cm]{fig1.eps}
%\caption{\label{label}
%\end{minipage}
%%\end{center}
%\end{figure}
%\hspace{2pc}%
\begin{figure}[h]
\begin{center}
%\begin{minipage}{14pc}
\includegraphics[width=12cm]{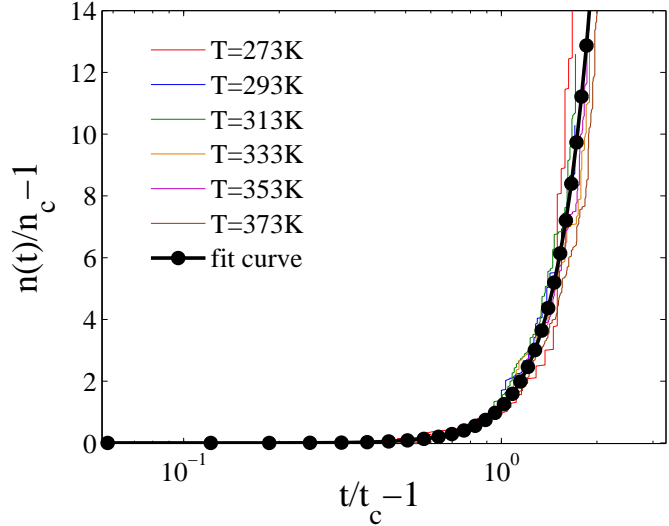}
\end{center}
\caption{\label{Fig_rescaled_growth_curves} Rescaled growth curves
at the different temperatures $T \in [273;\; 373]$~K (the pressure
$p=1$~atm); $t_c$ is the lag-time of the droplet with the critical
size $n_c$. Full circles correspond to the fit (see text for
details).}
%\end{minipage}
\end{figure}
The basic idea of the MFPT-approach is the next. First, it is
necessary to focus on the reaction coordinate and to define  the
time scales of the first appearance of its values, $\tau(n)$. As a
reaction coordinate for the case of the droplet nucleation, it is
convenient to utilize the droplet size $n$, which defines the
number of molecules (particles) involved in the nucleus of new
(liquid) phase. Second, by defining these time scales from $M$
independent `experiments' and by averaging results, one obtains
the MFPT-distribution:
\[
\tau^{\textrm{MFPT}}(n) = \sum_{i=1}^M \tau_i(n).
\]

In fact, the MFPT-distribution contains the information about
nucleation and nucleus growth kinetics. For example, the first
inflection points in the MFPT-curve defines the critical size
$n_c$ with the lag-time $t_c = \tau^{\mathrm{MFPT}}(n_c)$ (for
details, see \cite{Wedekind_JCP_2007}). Then, the part of the
inverted MFPT-distribution $n(\tau^{\textrm{MFPT}})|_{n \geq n_c}$
corresponds to the most probable growth law of the nucleated
droplet. Let us suppose the growth law  in the following
generalized form \cite{Mokshin/Barrat_PRE_2012}:
\begin{equation} \label{eq_growth_law_1}
n(t,t_c) = n_c + c_g \rho_{c}\mathcal{G}_{c}^{d\nu}(t -
t_c)^{d\nu},
\end{equation}
where $c_g$ is the shape-factor, $\rho_c$ is the density of the
nucleated phase, $\nu$ is the growth exponent -- it is equal to
unity in the case of constant growth rate, $\mathcal{G}_{c}$ is
the growth parameter and $d$ is the system dimension. Without
losing the generality in such a definition, the growth rate can be
time-dependent term and takes the next form : $G(t) = \nu
\mathcal{G}_{c}^{\nu} t^{\nu - 1}$ (see also
Ref.~\cite{Kashchiev_Nucleation}). By fitting
equation~(\ref{eq_growth_law_1}) to the distribution
$n(\tau^{\textrm{MFPT}})|_{n \geq n_c}$, one can extract the
values of the growth characteristics.

Molecular dynamics $NpT$-simulations were performed with the
coarse-grained mW-model~\cite{Molinero_JPCB_2009} of water vapor
at the temperatures $T=273$, $283$, $293$, $303$, $313$, $323$,
$333$, $343$, $353$, $363$ and $373$~K at the pressure
$p=1$~atm.\footnote{The initial configurations were obtained by
the isobaric cooling from the equilibrated vapor at the
temperature $T=900$~K and at the pressure $p=1$~atm.} For an each
$(p,T)$-point the set from $100$ independent runs at the identical
initial conditions was carried out, and the time evolutions of the
largest nucleated (liquid) droplet were stored for every run.
Identification of liquid clusters was performed on the basis of
the cluster analysis with the Stillinger
rule~\cite{Mokshin/Galimzyanov_2012,Stillinger_JCP_1963}.

Figure~\ref{fig_growth_curves} shows the growth curves of water
droplet at the different temperatures. These results are obtained
from molecular dynamics simulations within the MFTP-approach. The
MFPT-treatment of these results reveals the interesting features
of the nucleation-growth kinetics. The critical size $n_c(T)$ has
a weak decay with the temperature (for the considered temperature
range and the fixed pressure), that contradicts the classical
nucleation theory predictions~\cite{Kashchiev_Nucleation}. On
other hand, the lag-time $t_c$ increases with temperature from
$t_c(T=273\;\mathrm{K})=0.69\;\mathrm{ns}$ to
$t_c(T=373\;\mathrm{K})=1.56\;\mathrm{ns}$. The nucleated droplets
have shapes, which are closed to spherical ones (see
figure~\ref{fig_snapshots of droplet}). Further, the growth
parameter $\mathcal{G}_{c}$ decreases with temperature from
$\mathcal{G}_{c}(T=273\;\mathrm{K})=576\;\mathrm{m^{\nu}/s}$ to
$\mathcal{G}_{c}(T=373\;\mathrm{K})=242\;\mathrm{m^{\nu}/s}$,
whilst the growth exponent is independent on the temperature and
takes the value $\nu = 1.3$. This indicates on the next time
dependence of the growth rate: $G(t) \sim t^{0.3}$. Other result
is that the growth curves rescaled to the form
\[
\frac{n(t)}{n_c} -1 = \frac{c_g \rho_{c}(\mathcal{G}_{c}
t_c)^{3\nu}}{n_c}\left (\frac{t}{t_c}-1 \right )^{d\nu}
\]
are independent on the temperature and collapse onto a single
growth curve (see figure~\ref{Fig_rescaled_growth_curves}).

\section{Conclusions}

The following results are presented in this work.

(i) The treatment of the experimental data and of the simulation
results on the nucleation-growth kinetics can be performed on the
basis of the MFPT-approach, which allows one to extract the main
characteristics of the cluster growth process.

(ii) The growth of water droplets was studied on the basis of
coarse-grained molecular dynamics simulations for a wide range of
the temperature.

(iii) It is found that the rescaled growth law of water droplets
within the mW-model is unified at all the considered temperatures.

(iv) It is established that the droplet growth evolves with
acceleration and follows the power law.

\end{document}